\documentclass[epj]{svjour}
\usepackage[utf8]{inputenc}

\usepackage[switch]{lineno}
\usepackage{graphicx}
\usepackage{multirow}
\usepackage{comment}
\usepackage{enumitem}
\usepackage{afterpage}
\usepackage{sidecap}
\usepackage[numbers,sort&compress]{natbib}
\usepackage{dblfloatfix} %To enable figures at the bottom of page
\graphicspath{{../Figures/}{.}}
\newcommand{\fig}[1]{fig.~\ref{#1}}
\newcommand{\Fig}[1]{Figure~\ref{#1}}

\newcommand{\Sec}[1]{sec.~\ref{#1}}

\newcommand{\textmu}{$\mu$}
\begin{document}
%\linenumbers
%
\title{Background in $\gamma$-ray detectors and carbon beam tests in the Felsenkeller shallow-underground accelerator laboratory}
\author{
Tamás Szücs\inst{1}\thanks{t.szuecs@hzdr.de} \and
Daniel Bemmerer\inst{1}\thanks{d.bemmerer@hzdr.de} \and
Detlev Degering\inst{3}\and
Alexander Domula\inst{2} \and
Marcel Grieger\inst{1,2}\and
Felix Ludwig\inst{1,2}\and
Konrad Schmidt\inst{2}\and
Julia Steckling\inst{1,2}\and
Steffen Turkat\inst{2}\and
Kai Zuber\inst{2}
}
\institute{
Helmholtz-Zentrum Dresden-Rossendorf (HZDR), Dresden, Germany \and
Institut für Kern- und Teilchenphysik, Technische Universität Dresden, Dresden, Germany \and
VKTA – Strahlenschutz, Analytik \& Entsorgung Rossendorf, Dresden, Germany
 }
\date{\today}
\abstract{
The relevant interaction energies for astrophysical radiative capture reactions are very low, much below the repulsive Coulomb barrier. This leads to low cross sections, low counting rates in $\gamma$-ray detectors, and therefore the need to perform such experiments at ion accelerators placed in underground settings, shielded from cosmic rays.  
Here, the feasibility of such experiments in the new shallow-underground accelerator laboratory in tunnels VIII and IX of the Felsenkeller site in Dresden, Germany, is evaluated. 
To this end, the no-beam background in three different types of germanium detectors, i.e. a Euroball/Miniball triple cluster and two large monolithic detectors, is measured over periods of 26-66 days. The cosmic-ray induced background is found to be reduced by a factor of 500-2400, by the combined effects of, first, the 140 meters water equivalent overburden attenuating the cosmic muon flux by a factor of 40, and second, scintillation veto detectors gating out most of the remaining muon-induced effects. The new background data are compared to spectra taken with the same detectors at the Earth's surface and at other underground sites. 
Subsequently, the beam intensity from the cesium sputter ion source installed in Felsenkeller has been studied over periods of several hours. Based on the background and beam intensity data reported here, for the example of the $^{12}$C($\alpha$,$\gamma$)$^{16}$O reaction it is shown that highly sensitive experiments will be possible. 
 % end of PACS codes
} %end of abstract
\authorrunning{T. Sz\"ucs \it {\it et al.}}
\titlerunning{Background and carbon beam tests in the Felsenkeller shallow-underground accelerator laboratory}
\maketitle
\section{Introduction}
\label{sec:intro}

In experimental nuclear astrophysics, nuclear reactions that are naturally occurring in stellar environments are studied in the laboratory. For a given stellar temperature, the relevant energy range where the reaction cross section must be known is called the Gamow window \cite{Rolfs88-Book}.

For charged-particle induced reactions at energies inside this Gamow window, the cross section is very small, in the nanobarn \cite{Bemmerer06-PRL} to femtobarn \cite{Lemut06-PLB} range. In order to successfully measure such small radiative-capture cross sections, high target thickness, high beam intensity, and low background are needed. 

Since many years, the LUNA collaboration has been studying nuclear reactions inside or near the Gamow window, using a 0.4\,MV accelerator for $^1$H$^+$ and $^4$He$^+$ ions deep underground in the Gran Sasso laboratory, Italy  \cite{Broggini18-PPNP}. The LUNA site is shielded by 1400\,m of rock, equivalent to 3800\,m of water, reducing the cosmic-ray muon flux by six orders of magnitude. 

The $\gamma$-ray background at LUNA has been extensively characterized for different types of $\gamma$-ray detectors: For high-purity germanium (HPGe) detectors \cite{Bemmerer05-EPJA,Caciolli09-EPJA}, a composite HPGe with an external veto detector \cite{Szucs10-EPJA}, and for scintillation detectors made of bismuth germanate (BGO) \cite{Bemmerer05-EPJA,Boeltzig18-JPG} and lanthanum bromide (LaBr$_3$) \cite{Szucs12-EPJA}.

The science cases that have been addressed there include solar hydrogen burning \cite{Formicola04-PLB,Lemut06-PLB,Bemmerer06-PRL}, Big Bang nucleosynthesis \cite{Bemmerer06-PRL,Anders14-PRL}, and a number of higher hydrogen burning reactions \cite{Bemmerer09-JPG,Caciolli11-AA,Scott12-PRL,Strieder12-PLB,Cavanna15-PRL,Bemmerer18-EPL,Ferraro18-PRL,Bruno19-PLB}.

However, in order to build on the successes of LUNA \cite{Broggini18-PPNP} and also to study helium and carbon burning as well as the neutron sources for the astrophysical s-process, higher-energy underground ion accelerators have been called for \cite{NuPECC17-LRP}. 

Currently there are two such efforts ongoing in Europe. The first, called LUNA-MV, foresees the installation of a 3.5\,MV accelerator for hydrogen, helium, and carbon beams in the Gran Sasso underground laboratory, Italy \cite{Sen19-NIMB}. The LUNA-MV accelerator is scheduled to be installed in 2019, and the science case for the first five years mentions the $^{14}$N(p,$\gamma$)$^{15}$O, $^{22}$Ne($\alpha$,n)$^{25}$Mg, $^{13}$C($\alpha$,n)$^{16}$O, and $^{12}$C+$^{12}$C reactions \cite{Broggini19-NCA}.

%================================================================================================
\begin{figure*}[!b]
\center
  \includegraphics[width=0.8\linewidth]{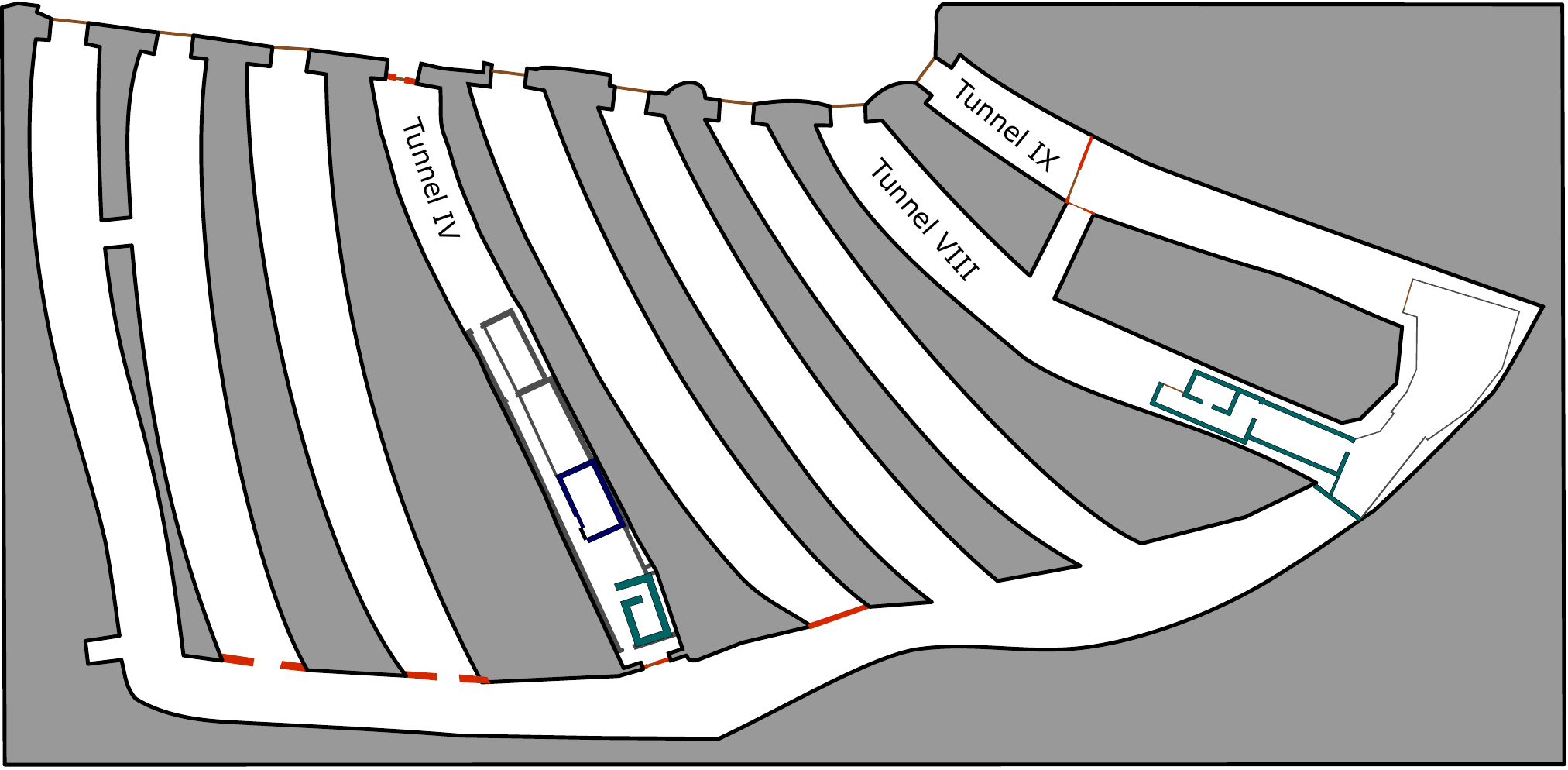}
\caption{The Felsenkeller tunnel system. The low-background activity-counting laboratory in tunnel IV was constructed in 1982 \cite{Helbig84-IIEHS} and enlarged in 1995 \cite{Niese96-Apradiso}. The data reported in the present work have been taken in the new laboratory \cite{Bemmerer18-SNC} in tunnels VIII and IX.}
\label{fig:Map_9_Tunnels}
\end{figure*}
%================================================================================================

The second project includes the installation of a 5\,MV accelerator in the Felsenkeller shallow-underground laboratory in Dresden, Germany \cite{Bemmerer18-SNC}. This site is shielded by 45 meters of rock (140 meters water equivalent, in short 140 m.w.e.) overburden \cite{Ludwig19-APP}. The civil construction of a new laboratory located in tunnels VIII and IX has been completed in 2018. The commissioning of the ion accelerator is ongoing. 

Outside of Europe, similar work is underway in the Sanford Underground Research Facility, USA \cite{Robertson16-EPJWOC} and at the Jinping laboratory, China \cite{Liu16-EPJWOC}.

For a successful scientific use of the Felsenkeller shallow-underground lab, both a low background in $\gamma$-ray detectors and a high ion beam intensity are needed. The present work aims to address these two aspects with dedicated experiments carried out in the new facility in Felsenkeller tunnels VIII and IX.

In order to address the first aspect, new $\gamma$-ray background data are reported, for three HPGe detectors that will actually serve for the planned experiments in Felsenkeller \cite{Bemmerer18-SNC}. 

These data build on two previous works \cite{Szucs12-EPJA,Szucs15-EPJA}. Ref. \cite{Szucs12-EPJA} showed a detailed comparison of the background in two sites, namely Felsenkeller (tunnel IV, 140 m.w.e.) and Gran Sasso (3800 m.w.e.). For the comparison, two detectors that are typical for nuclear astrophysics experiments, namely an escape-suppressed Clover-type HPGe detector and a LaBr$_3$ scintillator, were used, in turn, at each of the sites studied. It was found that the shallow-underground background, when muons were gated out by the escape suppression shield, was a factor of 2-4 higher than deep underground \cite{Szucs12-EPJA}.
Ref. \cite{Szucs15-EPJA} extended the comparison to a third underground site of intermediate rock cover, the Reiche Zeche mine in Freiberg/Germany (400 m.w.e.). The muon-vetoed background at the latter site was again slightly lower than at Felsenkeller \cite{Szucs15-EPJA}. 

The new $\gamma$-ray background data shown in the present work are in the context of two additional background measurements at Felsenkeller: One of the muon flux and angular distribution \cite{Ludwig19-APP}, and another of the neutron flux and energy spectrum \cite{Grieger19-PRD}. Taken together, these three studies \cite[and the present work]{Ludwig19-APP,Grieger19-PRD} serve to completely characterize this shallow-underground site. It is hoped that it can then be used for reference purposes in the future.

The second aspect needed for a successful scientific use of Felsenkeller, i.e. the ion beam intensity, is addressed here by tests of the carbon ion source that have been conducted in its final location and configuration in Felsenkeller tunnel IX. 

This work is organized as follows. In \Sec{sec:FK}, the new laboratory and its main equipments are presented, and \Sec{sec:bck} describes the setup of the $\gamma$-ray background studies. In \Sec{sec:res}, the measured $\gamma$-ray background is reported. An additional off-line $\gamma$-ray counting detector that was installed in the new laboratory is described in \Sec{sec:TU-1}. Sec.~\ref{sec:Ion_source} presents the measured beam current and stability of the negative ion source of the accelerator in its final configuration. An outlook is given in \Sec{sec:outlook}, and the present work is summarized in \Sec{sec:Summary}.

\section{The Felsenkeller accelerator laboratory}
\label{sec:FK}

The new Felsenkeller accelerator laboratory is situated in tunnels VIII and IX of the Felsenkeller tunnel system (\Fig{fig:Map_9_Tunnels}). The tunnels are part of an industrial area within the city limits of Dresden. Since 1982, a low level counting laboratory is operating in tunnel IV \cite{Helbig84-IIEHS,Niese96-Apradiso,Niese98-JRNC}. 

%================================================================================================
\begin{figure}[!b]
\center
  \includegraphics[width=0.85\linewidth]{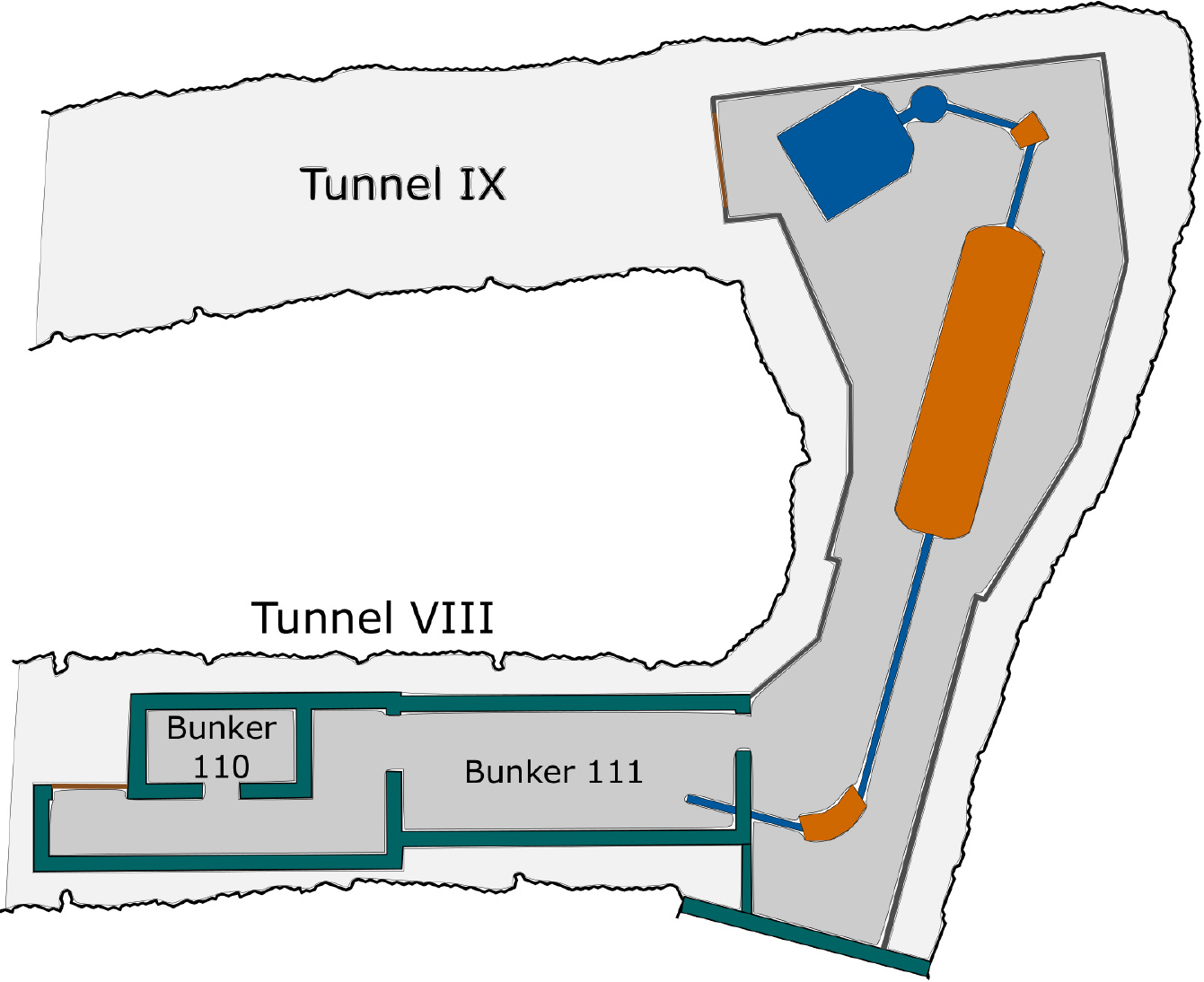}
\caption{Layout of the Felsenkeller accelerator laboratory in tunnels VIII and IX. The carbon sputter ion source (blue, in tunnel IX), the Pelletron accelerator and low and high energy magnets (orange) are shown. The air-conditioned area that is also radiation safety controlled area is shaded in dark grey. Walls by 40\,cm thick low-background concrete are shown as thick dark-green lines. See text for details.}
\label{fig:Map_TunnelsVIII_IX}
\end{figure}
%================================================================================================

\subsection{Description of the building}
\label{subsec:Building}

The entire Felsenkeller tunnel system has been created in 1856-1859 and was initially used as ice cellars for the Felsenkeller brewery. Tunnels VIII and IX had been left largely unchanged since that time. The surrounding rock consists mainly of hornblende monzonite \cite{Paelchen08-Book}. Samples taken at different places in the tunnels show specific activities of 130$\pm$30 and 170$\pm$30 Bq/kg for $^{238}$U and $^{232}$Th, respectively \cite{Grieger16-Master}. Before the beginning of the construction phase, extensive improvements were carried out on the walls and ceilings of tunnels VIII and IX, ensuring rock safety for the next 50 years. 

Subsequently, two experimental bunkers called 110 and 111, respectively, were installed in tunnel VIII (\Fig{fig:Map_TunnelsVIII_IX}). The bunkers are surrounded by 40\,cm thick reinforced concrete wall, floor, and ceiling. The materials (sand, gravel, cement, and ash) for the concrete were studied by $\gamma$-spec\-troscopy. For each lot of concrete, this study was carried out on the day before the mixing and pouring. The average specific activity was in the range of 14.9 - 17.8 and 15.6 - 17.4 Bq/kg for $^{238}$U and $^{232}$Th, respectively \cite{Bemmerer18-SNC}.

The concrete walls had been designed to limit two kinds of backgrounds: First, direct $\gamma$-rays emanating from the natural radioactivity of the surrounding rock, which are attenuated by the 40\,cm thick concrete. Second, neutron background produced in the walls by ($\mu$,n) reactions. In the concrete used here, the ($\mu$,n) yield is much lower than in commonly used high charge number shielding materials such as old iron or lead \cite{Grieger19-PRD}. 

The first of the two bunkers, called bunker 110, hosts several offline $\gamma$-counting setups (section \ref{sec:TU-1}).

The second bunker, called bunker 111, has been constructed for in-beam $\gamma$-ray spectroscopy experiments. It is connected to the accelerator hall by an open doorway and a hole in the bunker wall for the beam line. 

For the actual experimental setups and detectors in bunkers 110 and 111, in addition to the building walls tailor-made shielding will be used to further attenuate the background.

\subsection{Pelletron ion accelerator}

A 5\,MV Pelletron tandem accelerator produced by National Electrostatics (NEC), Wisconsin, USA, of type \linebreak 15SDH-2 is installed in Felsenkeller. This accelerator had been in operation from 1999 to 2012 in York/UK as a driver for $^{14}$C analyses at 4.5\,MV terminal voltage for the company Xceleron \cite{Young08-York}. In 2012, it was acquired by HZDR and transported to Dresden/Germany. 

The Pelletron accelerator has been installed in the very back of the connecting tunnel between tunnels VIII and IX (Figure \ref{fig:Map_TunnelsVIII_IX}). It has two different ion sources, see below. At a given time, it can run with one of these two sources.

The first ion source is an external cesium sputtering ion source of type 134 MC-SNICS, made by NEC, which is capable of delivering up to 100\,\textmu A $^{12}$C$^-$ beam \cite{NEC_sputter_source}. Test results obtained with this ion source in its final place of installation in the tunnel \cite{Steckling19-BSc} are shown below (Section \ref{sec:Ion_source}).    %or $^{1}$H$^-$ 

The second ion source is an internal radio frequency (RF) ion source made by NEC. It is mounted on the high-voltage terminal, behind a custom-made electrostatic deflector so that both tandem mode and single-ended mode operations are possible \cite{Reinicke18-PhD}. At a test setup at HZDR, the RF ion source delivered up to 90\,\textmu A $^{4}$He$^+$ beam. 

As a result, the Pelletron can be operated either in tandem mode for higher mass particles (for example C, N, O) or single ended mode for light ions (for example H, He) or noble gas ions. With this unique combination, a variety of hydrogen and helium burning reactions are planned to be investigated either in direct or in inverse kinematics.

%================================================================================================
\begin{figure*}[t]
\centering
  \includegraphics[clip, trim=5cm 1.5cm 5cm 1.5cm,width=0.3\linewidth]{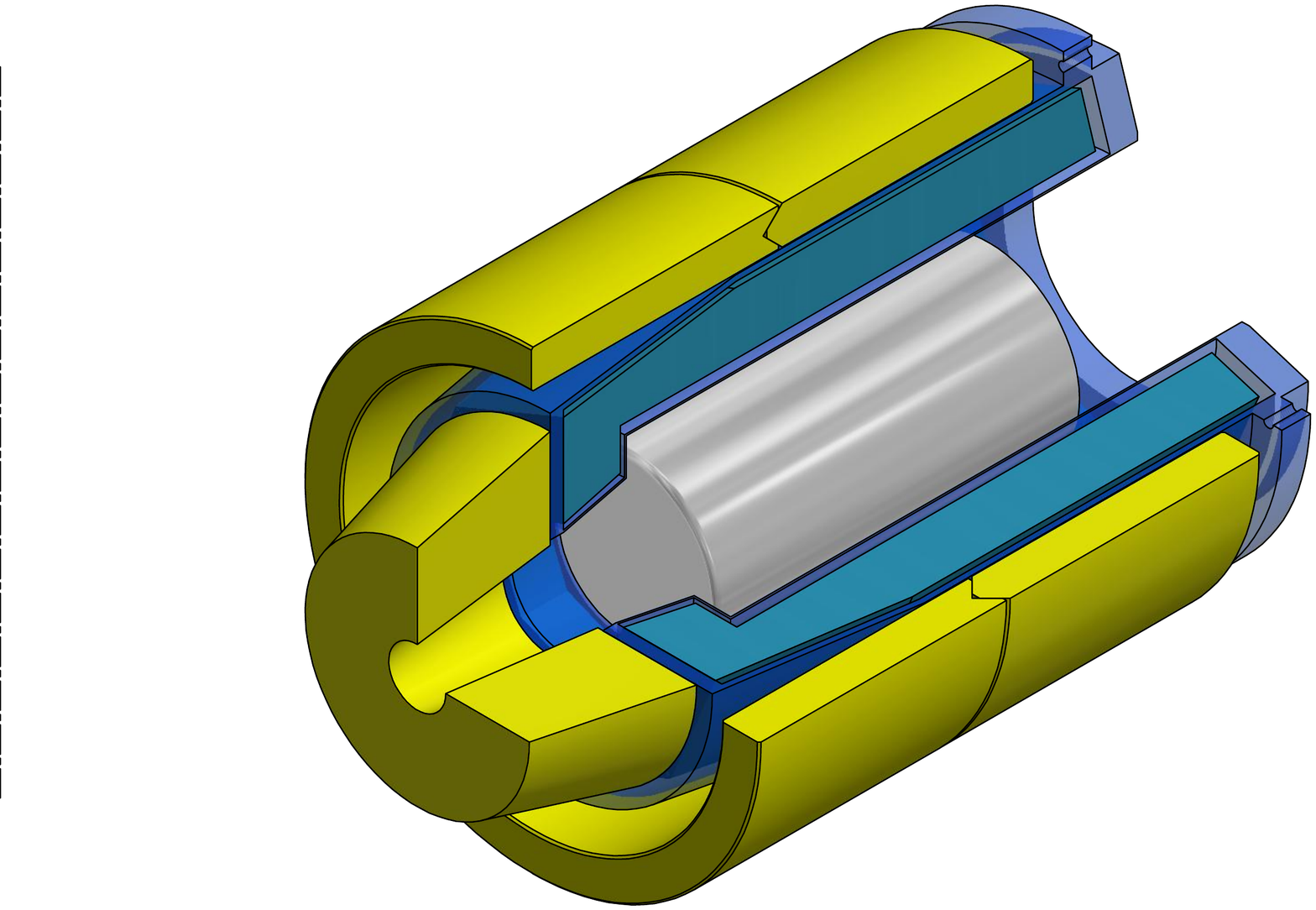}
  \includegraphics[clip, trim=5cm 1.5cm 5cm 1.5cm,width=0.3\linewidth]{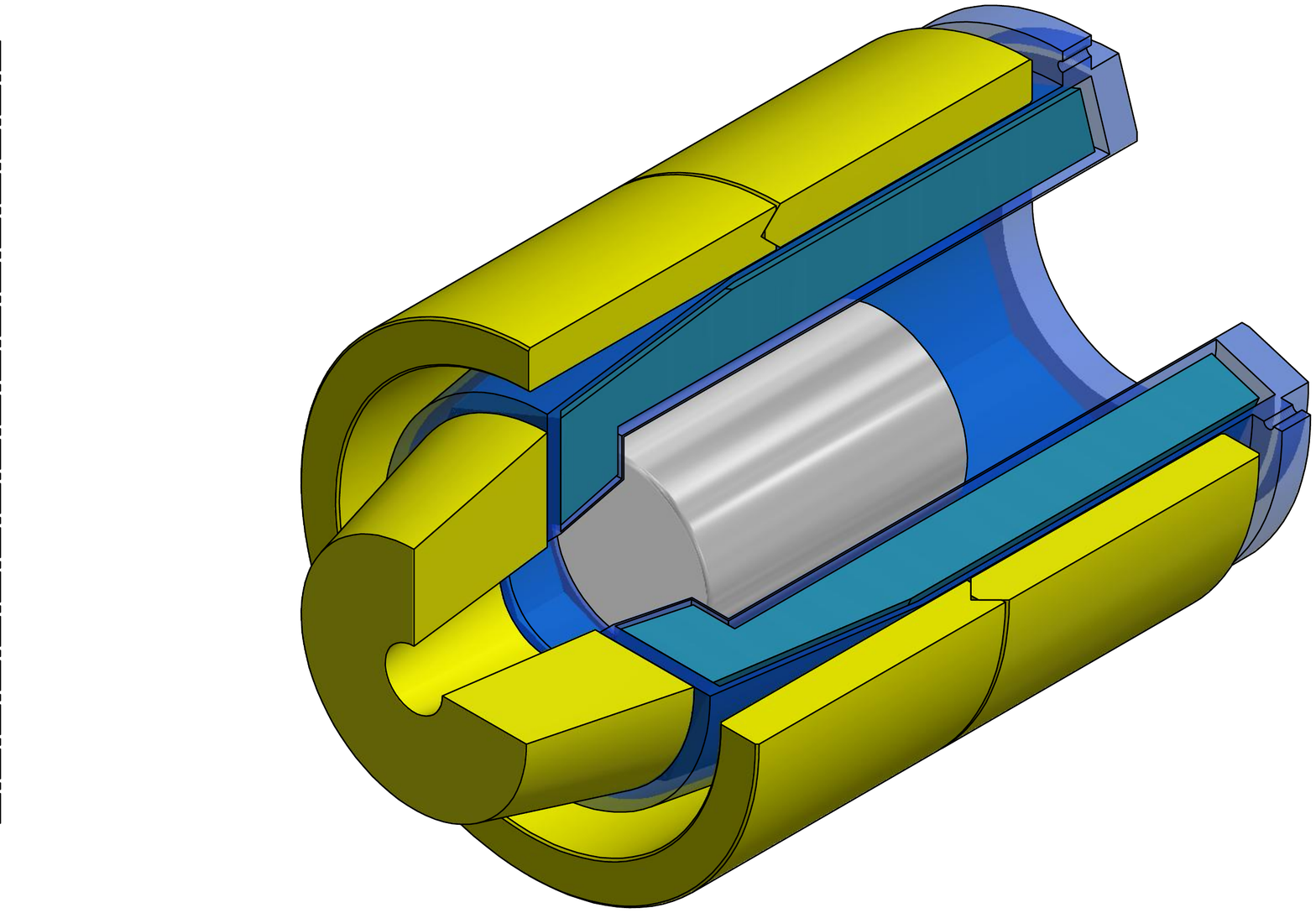}
  \includegraphics[clip, trim=6cm 1.5cm 5cm 1.5cm,width=0.3\linewidth]{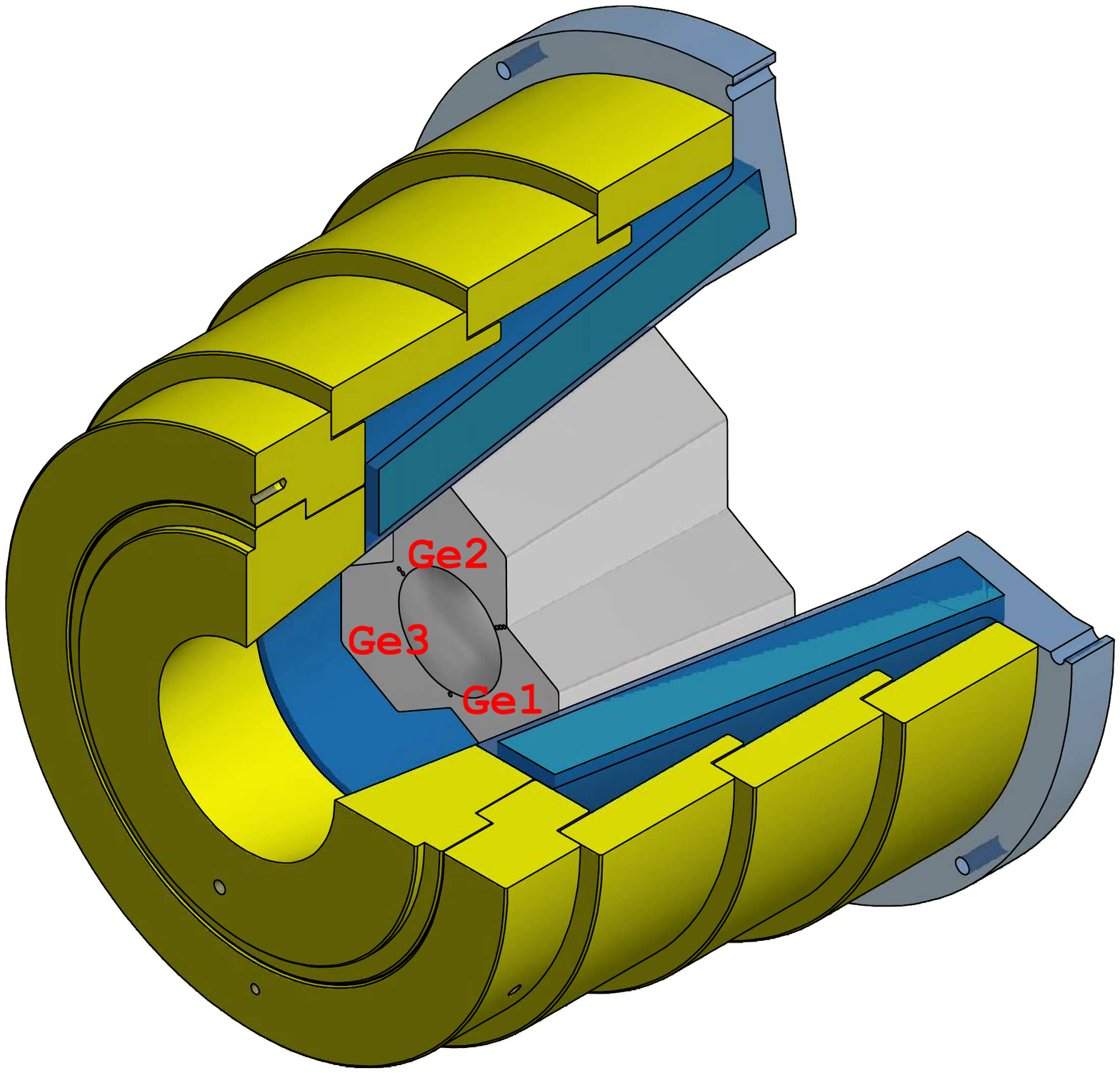}
\caption{Schematic view of the three $\gamma$-ray detectors: 90\% HPGe (called HZDR-1), 60\% HPGe (HZDR-2), and Euroball/Miniball (HZDR-3), from left to right. The end caps are shown in gray, the BGO veto detectors in blue and the lead shielding and collimators in yellow. For clarity, the necks of the HPGe detectors, the phototubes of the veto detectors and all holding structures have been removed from the picture.}
\label{fig:Schematic_HPGe_BGO}
\end{figure*}
%================================================================================================

\section{$\gamma$-ray background measurements with escape-suppressed HPGe detectors}
\label{sec:bck}

As a first step, the $\gamma$-ray background was measured with three typical escape-suppressed HPGe detectors, one of which had been used in a previous background intercomparison \cite{Szucs15-EPJA}. All three are foreseen to be used in future accelerator-based experiments at Felsenkeller. 

\subsection{$\gamma$-ray detectors used}

Detectors HZDR-1 and HZDR-2 are coaxial p-type HPGe detectors of cylindric shape: HZDR-1 with 88\% relative efficiency%\footnote{In this notation, 100\% relative efficiency corresponds to the detection efficiency of a 3''$\times$3'' NaI detector at 1.33 MeV $\gamma$-ray energy, with a $^{60}$Co source placed at 25\,cm distance, in the direction of the symmetry axis of the detector
~\cite{Gilmore08-Book}%.}
~(crystal diameter 80\,mm, length 78\,mm) and a carbon-fiber endcap, and HZDR-2 with 60\% relative efficiency (crystal diameter 71\,mm, length 60\,mm) and an aluminum endcap. 

Detector HZDR-1 has already been used for a number of experiments at the Earth's surface \cite{Reinicke18-PhD,Schmidt13-PRC,Schmidt14-PRC,Wagner18-PRC} and also deep underground \cite{Cavanna15-PRL,Depalo16-PRC}. The detector has been manufactured to ultra-low background specifications, and crystal and preamplifier are separated by a 25\,cm long neck, making it possible to mount this detector in a heavy lead shielding if needed.

Detector HZDR-2 has a 15\,cm long neck between crystal and preamplifier, but no special low-background configuration. Also this detector has been used for in-beam experiments at the Earth's surface \cite{Marta10-PRC,Schmidt13-PRC,Schmidt14-PRC,Wagner18-PRC,Reinicke18-PhD} and for a previous underground background measurement \cite{Szucs15-EPJA}.

Detector HZDR-3 consists of three Euroball HPGe detectors (60\% relative efficiency each, \cite{Wilhelm96-NIMA}), which are mounted together in a Miniball \cite{Reiter02-NPA} triple cryostat. For simplicity, this detector is called "Euroball/Miniball" here. When used in so-called add-back mode, i.e. with the energy deposited in all three crystals summed together, HZDR-3 has a relative efficiency of 240\%. 

Each of the three HPGe detectors HZDR-1, -2, and -3 is fitted with an escape suppression shield, which is realized in each case by a BGO scintillator surrounding the active germanium area (Figure \ref{fig:Schematic_HPGe_BGO}), in order to suppress events caused by the remaining muon flux \cite{Ludwig19-APP} in Felsenkeller. The BGO scintillator, in turn, is passively shielded against radiation from the sides (3\,cm of lead) and from the target area (7\,cm of lead). 

For detectors HZDR-1 and HZDR-2, the BGO shield has a thickness of 5\,cm, extends to 7\,cm (HZDR-1) or 10\,cm (HZDR-2) behind the back end of the crystal and includes a front shield that is 5\,cm thick, with an opening of 6\,cm diameter that limits the effective area of the detector. For the case of detector HZDR-3, the BGO shield is just 3\,cm thick and extends to 5\,cm beyond the front of the crystal and 5\,cm beyond the back end of the crystal (see \fig{fig:Schematic_HPGe_BGO}).

\subsection{Data acquisition and offline data sorting}
\label{subsec:DAQ}

The signals from both HPGe and BGO detectors were recorded as time-stamped energy data, and coincident and anti-coincident events were reconstructed offline after the end of the measurement.

The data have been recorded with a CAEN V1725 digitizer with 250 MS/s sampling rate and DPP-PHA firmware (version 128) in list mode, with every channel internally self-triggered. For the runs analyzed here, the least significant bit in time was  4\,ns, and the energy signals were histogrammed in 16,384 channels. For the sake of simplicity, higher possible resolutions were not used.

In the offline analysis, the timestamped data were sorted into events, and histograms with and without muon veto were built. 
As expected, no suppression in the total peak areas of the radionuclide background peaks  was found due to the BGO veto. 

\subsection{Measurement procedure}
\label{subsec:Measurements}

%================================================================================================
\begin{figure}[b]
\center
  \includegraphics[width=0.95\linewidth]{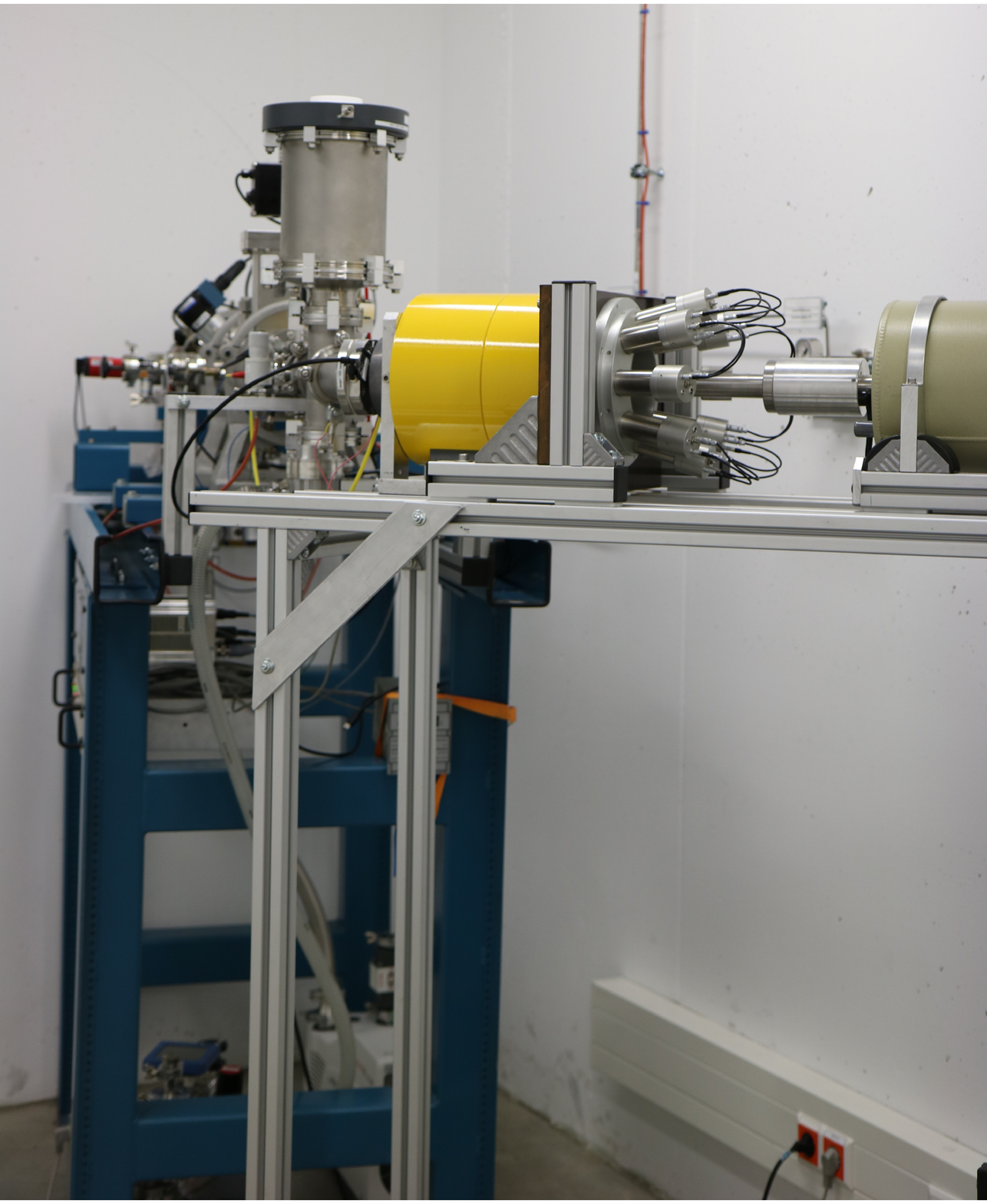}
\caption{Photograph of a test irradiation chamber, connected to the Pelletron beam line, and the escape-suppressed HPGe detector HZDR-1 in tunnel VIII, bunker 111. The liquid nitrogen dewar for the cold trap is seen atop the target chamber. The lead shield surrounding the BGO escape suppression of the HPGe detector is seen in yellow. The beam level is 160\,cm above the concrete floor.}
\label{fig:Photo_Targetchamber}
\end{figure}
%================================================================================================

For the underground background measurements, each of the detectors HZDR-1, -2, and -3 was, in turn, mounted horizontally on an aluminum stand. The symmetry axis of the detector was 160\,cm above the floor (Figure \ref{fig:Photo_Targetchamber}). The detectors were placed in Felsenkeller tunnel VIII, bunker 111. 

The running times in the underground laboratory were 26, 36, and 66 days, respectively, for detectors HZDR-1, -2, and -3.

For reference, also overground data were taken at the HZDR Rossendorf site (300 m above sea level). For detectors HZDR-1 and HZDR-3, the overground spectra were taken in the one-story 3\,MV Tandetron building. For HZDR-2, the overground spectrum was taken in the basement of a three-story building.

%================================================================================================

\section{Results of the measurements with escape-suppressed HPGe detectors}
\label{sec:res}

In the following, the data from the three detectors HZDR-1, -2, and -3 are discussed, and then the general features common to all three detectors are summarised. 

It is noted that for much of this section, the discussion is limited to $\gamma$-ray energies $E_\gamma$ $>$ 3.5\,MeV, where the radionuclide background plays no direct role. When very low background counting rates are aimed at for the radionuclide-dominated energy range $E_\gamma$ $\leq$ 3.5\,MeV, in addition to passive and active muon suppression measures, also elaborate passive lead and copper shields are required, see section \ref{sec:TU-1} below.

\subsection{Detector HZDR-1: 90\%\,HPGe}
\label{sec:90}

\begin{figure}[b]
\center
  \includegraphics{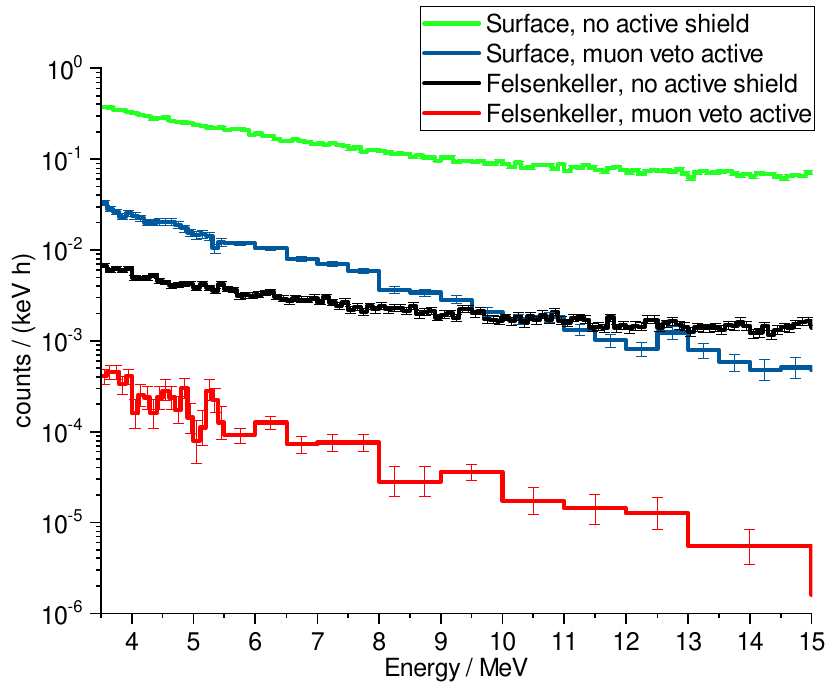}
\caption{$\gamma$-ray energy spectra recorded with detector HZDR-1 (90\%\,HPGe) at the Earth's surface (without and with muon veto) and underground at Felsenkeller, tunnel VIII, room 111 (with and without muon veto). See text for details.}
\label{fig:90HPGe}
\end{figure}

In detector HZDR-1, the no-beam background spectra do not exhibit any marked features at energies $E_\gamma$ $\geq$ 3.5\,MeV, just a general decrease with energy that is characteristic of cosmic-ray induced effects (Figure \ref{fig:90HPGe}). As expected, this continuum is generally decreased, by a factor of 30-100, when comparing overground and Felsenkeller spectra. In this detector, the muon peak is expected at $E_\gamma \approx$ 60\,MeV \cite{Szucs15-EPJA}, outside the scope of Figure \ref{fig:90HPGe}. 

%================================================================================================
%================================================================================================

The detailed background counting rates in several energy regions, which were chosen to match those from a previous study \cite{Szucs15-EPJA}, are shown in Table \ref{Table:HZDR1}. The first energy region, $E_\gamma$ = 6-8 MeV, starts above the energy where $\alpha$ activity can be present in the soldering of the crystal \cite{Brodzinski87-NIMA}. At all displayed energies, the penetrating muons dominate the background at the surface of the earth. However, with increasing depth, the relative importance of neutrons also increases. 

%================================================================================================
\begin{table*}[t]
\caption{Recorded counting rates in detector HZDR-1 (90\% HPGe) for different energy regions, in units of 10$^{-3}$ counts / (keV hour). For the 6--8\,MeV energy range, for comparison literature data \cite{Szucs10-EPJA,Best16-EPJA} from two larger HPGe detectors used deep underground at Gran Sasso are also shown.}
\label{Table:HZDR1}
\center
\begin{tabular}{l|r@{\,}c@{\,}l r@{\,}c@{\,}l r@{\,}c@{\,}l}
\noalign{\smallskip}\hline\noalign{\smallskip}
Site						&	\multicolumn{3}{c}{6--8\,MeV}	&	\multicolumn{3}{c}{8--10\,MeV}	&	\multicolumn{3}{c}{10--15\,MeV}	\\
\noalign{\smallskip}\hline\noalign{\smallskip}
Surface, no veto	&	151.7	&	$\pm$	&	1.0		&	103.4	&	$\pm$	&	0.9		&	74.9	&	$\pm$	&	0.5		\\
Surface, with veto			&	7.8		&	$\pm$	&	0.2		&	2.98		&	$\pm$	&	0.14	&	1.03	&	$\pm$	&	0.05	\\
Felsenkeller, no veto	&	2.75	&	$\pm$	&	0.05	&	2.07	&	$\pm$	&	0.04	&	1.54	&	$\pm$	&	0.02	\\
Felsenkeller, with veto	&	0.088	&	$\pm$	&	0.008	&	0.032	&	$\pm$	&	0.005	&	0.0111	&	$\pm$	&	0.0019 \\
\noalign{\smallskip}\hline\noalign{\smallskip}		

Gran Sasso, no veto, Clover 122\% \cite{Szucs10-EPJA}	&	0.15	&	$\pm$	&	0.03	\\
Gran Sasso, no veto, HPGe 136\% \cite{Best16-EPJA}	&	0.015	&	$\pm$	&	0.006	&	0.003	&	$\pm$	&	0.002 &	\multicolumn{3}{c}{$< 0.0011$}\\
\noalign{\smallskip}\hline										
\end{tabular}												
\end{table*}
%================================================================================================

At $E_\gamma$ = 6-8 MeV, also reference data from the deep-underground Gran Sasso laboratory are available \cite{Szucs10-EPJA,Best16-EPJA}, which have been taken using somewhat larger detectors. When scaling up the present background data to correct for the lower detector volume, the present background with muon veto is 0.8 times that of the 122\% Clover in Gran Sasso \cite{Szucs10-EPJA}, which was, just as the present detector HZDR-1, shielded by a BGO escape-suppression shield. Taking the uncertainties involved in comparing different detectors into account, these background values are mutually consistent (Table \ref{Table:HZDR1}).
It is noted that due to the extremely low muon background in Gran Sasso, the escape-suppression shield did not reduce the background there \cite{Szucs10-EPJA}.

%================================================================================================
\begin{figure*}[b]
\center
  \includegraphics{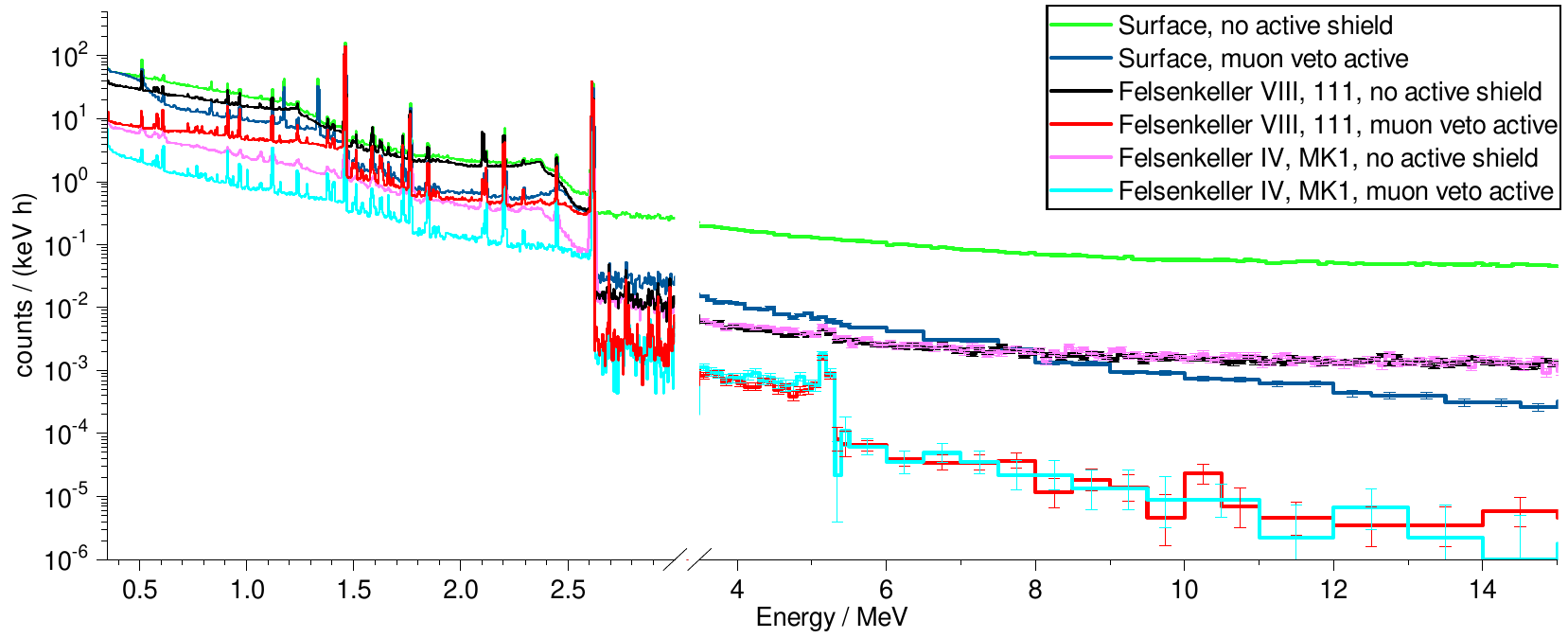}
\caption{$\gamma$-ray energy spectra recorded with detector HZDR-2 (60\%\,HPGe) at the Earth's surface, underground at Felsenkeller, tunnel VIII, room 111, and at Felsenkeller tunnel IV, MK1. See text for details.}
\label{fig:60HPGe}
\end{figure*}
%================================================================================================

Again taking the detector size into account, the present background with muon veto is 9 times that of a 136\% HPGe detector used at Gran Sasso, as well \cite{Best16-EPJA}. In ref. \cite{Best16-EPJA}, the HPGe detector was surrounded by a heavy passive shielding of 20\,cm lead with an inner liner of 5\,cm oxygen free high conductivity copper \cite{Cavanna14-EPJA}. This shield may have attenuated the ambient neutron flux, contributing to the lower background when compared to the 122\% Clover in Gran Sasso \cite{Szucs10-EPJA}.

At higher energies, 8-10 and 10-15 MeV, the present background counting rates drop quickly when compared to lower energies (Table \ref{Table:HZDR1}). Deep underground, the background drops even more rapidly \cite{Best16-EPJA}, as expected also based on studies with other types of detectors \cite{Bemmerer05-EPJA,Szucs12-EPJA}.  

When comparing the HZDR-1 counting rates with and without muon veto above and below ground, it is clear that the  cosmic-ray muon veto by the BGO detector becomes more effective underground: Whereas overground, the continuum for $E_\gamma \geq$ 6\,MeV is reduced by a factor of 20-70 by the muon veto, underground the reduction is a factor of 30-140. This result is consistent with previous work using similar detectors \cite{Szucs12-EPJA,Szucs15-EPJA}.

By combining the background reduction by the Felsenkeller rock overburden with the muon veto, for the characteristic 6-8\,MeV energy range in detector HZDR-1 a background reduction of a factor of 1730$\pm$160 is found.

%================================================================================================
\subsection{Detector HZDR-2: 60\%\,HPGe}
\label{sec:60}

For detector HZDR-2, which has been used previously in Felsenkeller tunnel IV and in the Reiche Zeche underground facility in Freiberg/Sachsen, Germany \cite{Szucs15-EPJA}, the background is studied from $E_\gamma$ = 0.4-15\,MeV here (Figure \ref{fig:60HPGe}).
 
At low energy, $E_\gamma$ $<$ 3\,MeV, the observed background is given by the radionuclides in the walls of the laboratory. In the new laboratory (Felsenkeller tunnel VIII, room 111), the specific activities for $^{232}$Th and $^{238}$U in the walls are in the 15-18 Bq/kg range (section \ref{subsec:Building}), lower than the original rock (130-170 Bq/kg) but higher than Felsenkeller tunnel IV, MK1: The serpentinite rock used there contains just 1.3 and 0.34\,Bq/kg of $^{238}$U and $^{232}$Th, respectively \cite{Niese98-JRNC}.

As a consequence, both the background lines and the continuum in Felsenkeller VIII/111 are lower then at the Earth's surface but higher then in Felsenkeller IV/MK1 (Figure \ref{fig:60HPGe}). Also the $^{40}$K line and its Compton continuum scale approximately as those of the uranium and thorium. 

The high-energy background, $E_\gamma$ $\geq$ 3.5\,MeV, behaves similarly to detector HZDR-1. No significant differences between the two Felsenkeller tunnels studied are visible with this detector, neither with nor without muon veto (Figure \ref{fig:60HPGe}). The main difference to HZDR-1 is that for the case of HZDR-2, an $\alpha$-induced peak at $E$ = 5.2\,MeV due to intrinsic impurities in the soldering \cite{Brodzinski87-NIMA} is clearly seen in the muon vetoed spectra. This is expected for a detector such as HZDR-2 where, unlike HZDR-1, no special low-background precautions have been taken in the manufacturing process.

Above ground the high energy ($E_\gamma \geq$ 6 MeV) background suppression by the muon veto is a factor of 30--100 in detector HZDR-2. In Felsenkeller tunnels VIII/111 and IV/MK1, it is a factor of 60--200.

For detector HZDR-2 and the characteristic 6-8\,MeV energy range, the combined effects of the Felsenkeller rock and the muon veto reduce the background by a factor of 2400$\pm$300.

\subsection{Detector HZDR-3: Euroball/Miniball}
\label{subsec:Miniball}
%================================================================================================
\begin{figure}[b]
\center
  \includegraphics{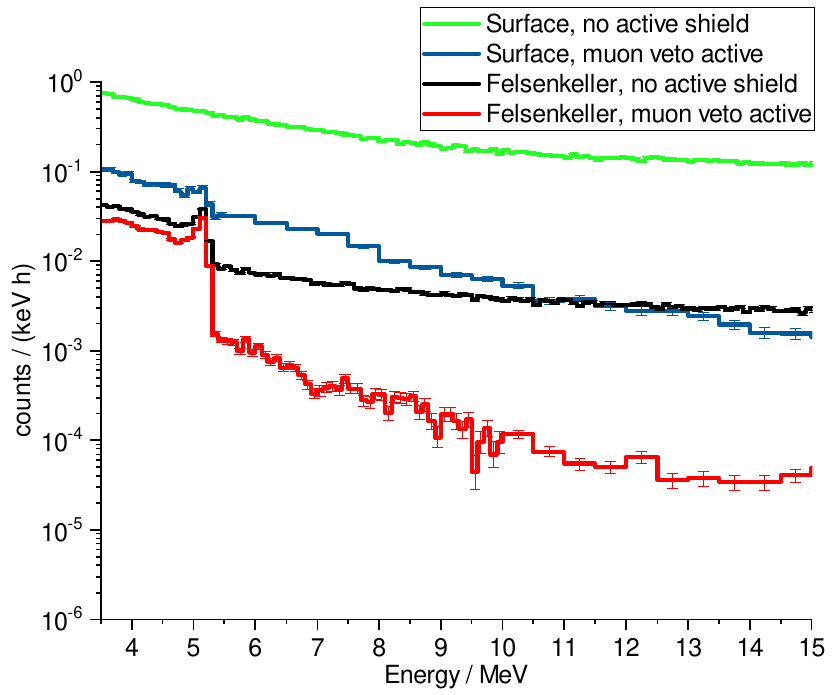}
\caption{Detector HZDR-3 (Euroball/Miniball), add-back mode spectra. See text for details.}
\label{fig:Addback}
\end{figure}
%================================================================================================
%================================================================================================
\begin{figure*}[t]
\center
  \includegraphics{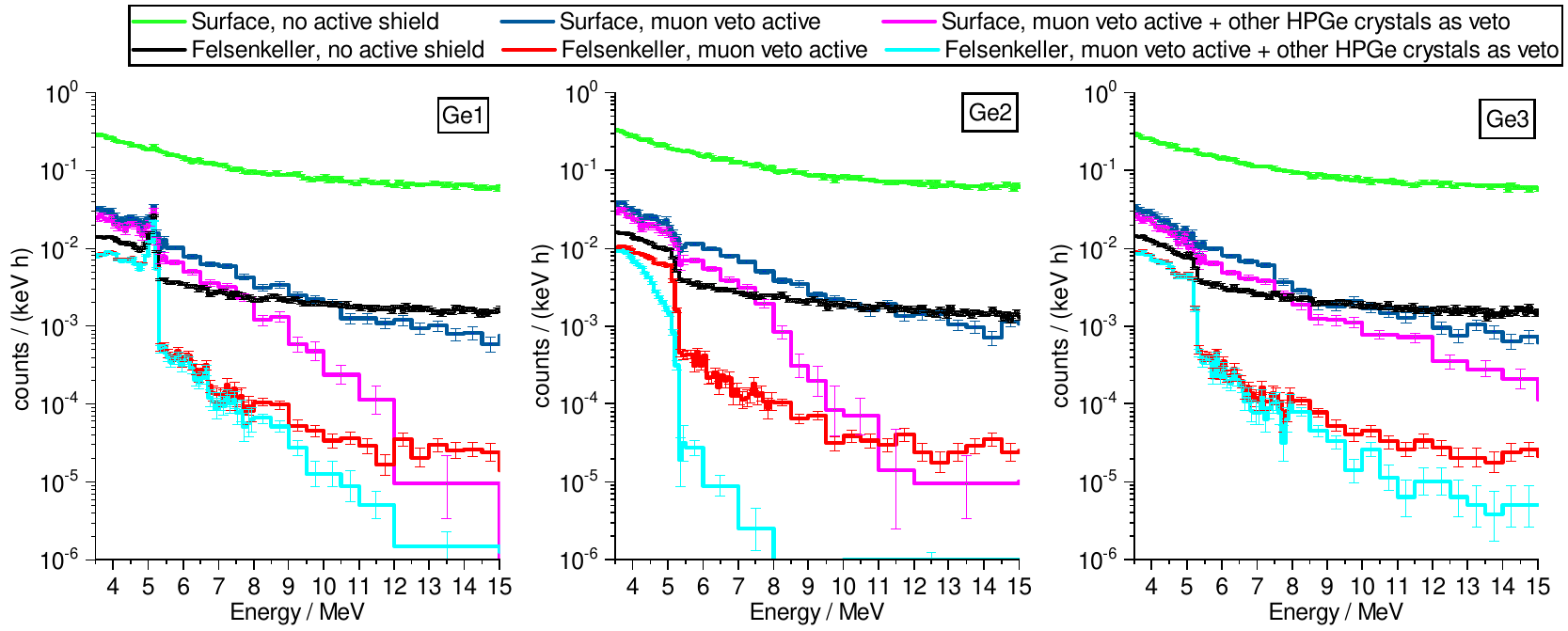}
\caption{Detector HZDR-3 (Euroball/Miniball), single spectra. See text for details.} 
\label{fig:Singles}
\end{figure*}
%================================================================================================

For detector HZDR-3, two types of spectra are available: First, the simple sum of the energy-calibrated histograms from the three crystals, hereafter called "singles mode". Second, the event-by-event sum of the entire energy deposited in any of the three Ge crystals is histogrammed as "add-back mode". The latter spectrum corresponds to a virtual large detector of 240\% relative efficiency. For both singles and add-back mode spectra, data with and without muon veto are available. 

The add-back mode spectra of detector HZDR-3 (Figure \ref{fig:Addback}) show the same general features as for detectors HZDR-1 and -2. As was the case for HZDR-2, there is again an $\alpha$-induced peak at $E_\gamma$ $\sim$ 5\,MeV. Above ground, for $E_\gamma >$ 6 MeV, the background suppression by the muon veto is a factor of 14 -- 50 in detector HZDR-3, add-back mode. Underground in Felsenkeller, it is a factor of 18 -- 60. 

When examining the single spectra of the three individual crystals (Figure \ref{fig:Singles}), some interesting effects are apparent. First, Ge1 has a much higher $\alpha$ contamination peak at 5 MeV than the other two crystals. Second, when using not only the BGO but also the other two Ge crystals as further muon veto, the count rate is again much suppressed for crystal Ge2, while there is almost no effect on Ge1 and Ge3. This may be explained by the fact that Ge2 is located atop Ge1 and Ge3 (figure \ref{fig:Schematic_HPGe_BGO}), so that most muons passing Ge2 register in one of the two lower Ge detectors. Conversely, there are many additional muons that pass either Ge1 or Ge3 but not Ge2. 

For the characteristic 6-8\,MeV energy range, the combined effects of the Felsenkeller rock and the muon veto reduce the HZDR-3 add-back mode background by a factor of 557$\pm$14. In single mode, the same factor is 688$\pm$17. 

Currently a second detector equal to HZDR-3 is in preparation and expected to come online in the fall of 2019.

%================================================================================================
\subsection{Comparison of the data from the all three detectors HZDR-1, HZDR-2, and HZDR-3}

Now the 6-8\,MeV counting rate suppression factors, which are given by the combined effects of the Felsenkeller site and the muon veto, can be compared for the three detectors HZDR-1, HZDR-2, and HZDR-3. 

The highest suppression factor, 2400$\pm$300, is found for HZDR-2, the smallest Ge crystal with the largest BGO veto detector  (Figure \ref{fig:Schematic_HPGe_BGO}). This may be compared to the HZDR-3 (single mode) value of 688$\pm$17, with a similar size of the single Ge crystal but a somewhat smaller BGO veto detector. So a larger BGO size (more precisely, a larger solid angle covered but partly also a greater BGO thickness) leads to a better muon suppression factor.

When comparing detectors HZDR-1 and HZDR-2, with the same BGO veto but a larger Ge crystal for HZDR-1, it seems that larger Ge crystals lead to lower suppression factors: For HZDR-1, it is just 1730$\pm$160.

In the case of HZDR-3, add-back mode, both large (virtual) Ge detector and smaller BGO veto lead to the lowest suppression factor observed, 557$\pm$14. For HZDR-3, the BGO veto covers a smaller solid angle of the Ge crystals than for the other two detectors (Figure \ref{fig:Schematic_HPGe_BGO}): The BGO inner diameter opens towards the back of the detector, and there is no part of the BGO covering part of the front of the Ge crystal.

Summarizing the absolute counting rates in the characteristic 6-8\,MeV energy region, for detectors HZDR-1 and HZDR-2, rates of just (0.088$\pm$0.008) $\times10^{-3}$\,keV$^{-1}$h$^{-1}$ and %\linebreak 
(0.036$\pm$0.005) $\times10^{-3}$\,keV$^{-1}$h$^{-1}$ are found, whereas for the HZDR-3 add-back mode the background is one order of magnitude higher, (0.525$\pm$0.013)$\times10^{-3}$\,keV$^{-1}$h$^{-1}$.

\section{Off-line $\gamma$-counting detector TU-1}
\label{sec:TU-1}

\subsection{Description of detector TU-1}

The off-line activity-counting bunker 110 (Figure \ref{fig:Map_TunnelsVIII_IX}) was designed to host several low-background detectors for activity determination purposes, both for nuclear astrophysics and also for general nuclear applications like decay branching ratio studies. 

The first detector installed there, here called TU-1, is a very large HPGe detector, with 574\,cm$^3$ active volume and 163\% relative efficiency, made by Canberra (Lingolsheim, France) to ultra-low-background specifications. The crystal is 90\,mm long and has 90\,mm diameter, with an initial dead layer of less than 0.5\,mm thickness. The measured resolution, full width at half maximum, of the 1.33\,MeV $^{60}$Co line is 1.92\,keV.  

The 20\,cm thick shielding is built up from three subsequent layers: The innermost layer is 5\,cm oxygen free radio-pure (OFRP) copper, specified to $\leq$0.04\,Bq/kg $^{238}$U (Aurubis). Subsequently, there is a 5\,cm thick lead layer, with $2.50\,\pm\,0.95$\,Bq/kg $^{210}$Pb (von Gahlen). The outermost layer is 10\,cm of lead with $21\,\pm\,2$\,Bq/kg $^{210}$Pb (R\"{o}hr und Stolberg). Outside the lead shielding, there is a provisoric anti-radon box made of plastic foil that is flushed by the evaporated nitrogen from the detector cryostat.

Inside this shield, there is a cylinder of 20\,cm diameter and 40\,cm height that contains the Ge detector end cap (115\,mm diameter, 196\,mm height) and space for large samples. The lid of the shielding can be removed by an electrically operated crane. 

The final configuration of detector TU-1 will include a plexiglass (PMME) anti-radon box and an active muon veto built up from 100$\times$100\,cm$^2$ plastic scintillator panels. In addition, it will be possible to add an inner liner of radiopure copper to partially fill the large inner area of the shield. The data reported here therefore represent an upper limit on the background to be expected in the future.

\subsection{Background data from detector TU-1, preliminary configuration}

%================================================================================================
\begin{figure}[b]
\center
  \includegraphics{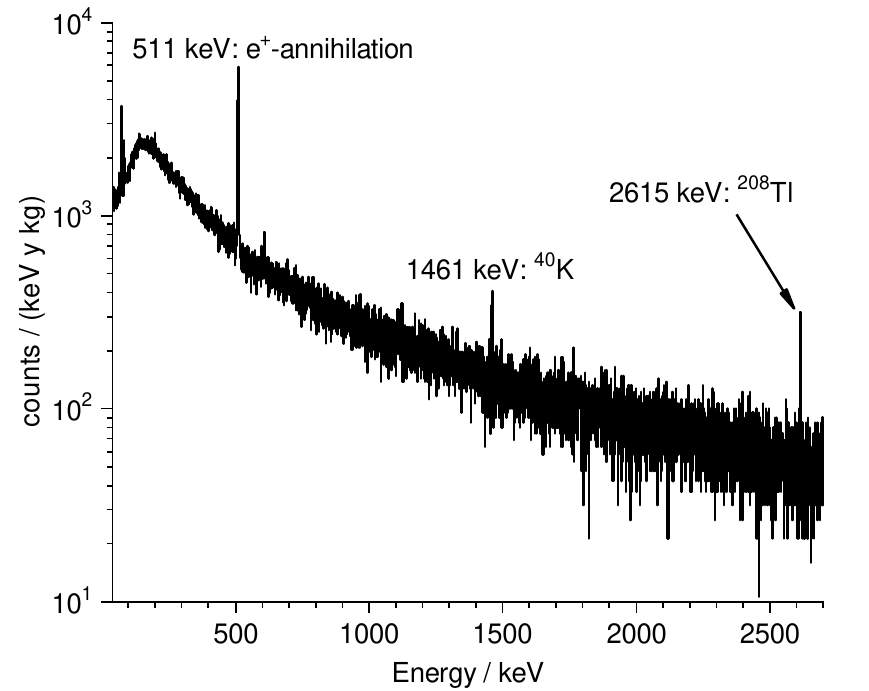}
   \caption{Detector TU-1, 49 day long background spectrum in its preliminary configuration. See text for details.}
\label{fig:GX150}
\end{figure}
%================================================================================================

In its preliminary configuration, i.e. full lead and copper shield, no inner liner, provisoric anti-radon box, and no anti-muon veto, a 49\,d long background run was performed with detector TU-1 (Figure \ref{fig:GX150}). The spectrum shows the $^{40}$K line at 1461\,keV, the $^{208}$Tl line at 2615\,keV, and a strong 511\,keV annihilation peak. In addition, radon-induced lines at 238 and 609 keV are in evidence. 

A further reduction in the counting rate of the $^{208}$Tl line is expected when the anti-radon box will be finalized. In addition, the planned anti-muon veto is expected to reduce the counting rates for both the 511\,keV line and the continuum. The source of the remaining $^{40}$K counts is being investigated while the setup is still being completed.

The integral activity, in this preliminary configuration, for $E_\gamma$ = 40-2700 keV is $(3043 \pm 5)$\,d$^{-1}$kg$^{-1}$. This is comparable to the value of 2938\,d$^{-1}$kg$^{-1}$ found previously in Felsenkeller tunnel IV for a similarly well-shielded HPGe detector of lower crystal size \cite{Kohler09-ARI}, but higher than in deep-underground laboratories \cite{Laubenstein04-ARI,Caciolli09-EPJA}. 

\section{Carbon beam tests of the sputter ion source}
\label{sec:Ion_source}

\subsection{Description of the sputter ion source}

The present cesium sputter ion source \cite{Middleton83-NIM} of type 134 MC-SNICS has been used previously from 1999-2012 at Xceleron in York/UK, for routine $^{14}$C analyses. It is a commercially available model manufactured by National Electrostatics (Middleton, Wisconsin, USA) and includes a cathode wheel capable of holding 134 samples. After being transported to Dresden, the 134 MC-SNICS was initially tested at low extraction voltage overground \cite{Koppitz17-BSc}.

In 2018, the source was brought to its final place of installation, underground in Felsenkeller tunnel IX (Figure \ref{fig:Map_TunnelsVIII_IX}). The control software was updated to a Labview-based system with Beckhoff hardware. Subsequently, commissioning and tests were carried out at 60\,kV total acceleration potential \cite{Steckling19-BSc}. 

\subsection{Long-term current measurement}

As part of the underground commissioning efforts, several different cathode materials to hold the graphite rods of 2\,mm diameter that are used for the production of $^{12}$C$^-$ ion beam were tested for their long-term viability. To this end, the accelerated $^{12}$C$^-$ beam was analysed by mass in the injector beam line electromagnet, in order to exclude contaminations by carbon molecules, cesium compounds, and the cathode holder materials. 

After an initial burn-in phase at 5\,kV cathode voltage, the cathode voltage was increased to 6\,kV when the observed current reached 40\,$\mu$A. This point was reached after 1.4 and 6.5 hours for the aluminum and copper cathode holder, respectively. When the cathode voltage was increased, the bias voltage was decreased by 1\,kV, so that the total beam energy, which is given by the sum of cathode, extraction, and bias voltages, remained constant at 60\,keV.

\begin{figure}[t]
\center
  \includegraphics{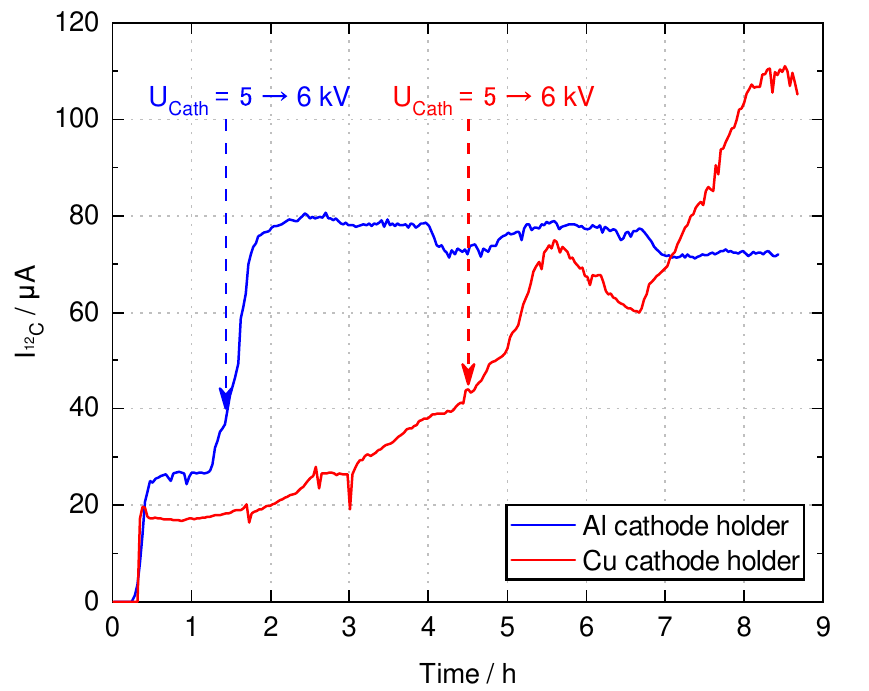}
\caption{One minute averaged, analyzed $^{12}$C$^-$ beam current from the Felsenkeller MC-SNICS ion source for two different cathode holder materials, as measured in the Faraday cup after the low-energy injection magnet.}
\label{fig:Current}
\end{figure}

The data show that the analysed beam has satisfactory intensity in the tens of \textmu A over at least 8 hours of operation (Figure \ref{fig:Current}). The aluminum cathode holder showed over 6 hours of stable operation in the 70-80\,\textmu A intensity range for analysed $^{12}$C$^-$ beam. For the copper cathode holder, the peak intensity was higher, but instead of a stable plateau current, there was a slow rise from 18 to 110 \textmu A, less well-suited for long-time irradiations.

%================================================================================================
%================================================================================================

When assuming the 6-hour interval that was apparent for aluminum cathode holders and taking into account that the cathode wheel can hold up to 134 different cathodes, it seems possible to operate the source continuously for one month without breaking vacuum. 

The aimed intensity for positively charged $^{12}$C beam after the accelerator is 50 particle-$\mu$A. When assuming a 30\% fraction for the selected charge state (typically $^{12}$C$^{+}$ or $^{12}$C$^{2+}$ for nuclear astrophysics experiments) out of the gas stripper installed on the terminal, the 70-80\,$\mu$A out of the Al cathode corresponds to 21-24\,particle-$\mu$A. As a consequence, for the final full performance, still a factor of two improvement over the initial performance shown here is needed.

\section{Outlook on planned future experiments}
\label{sec:outlook}

In stellar helium burning, the $^{12}$C($\alpha,\gamma$)$^{16}$O reaction has wide-ranging effects on the nucleosynthetic output \cite{Woosley07-PhysRep}. Its precise astrophysical S-factor at Gamow energies, roughly  $E$ = 0.2-0.6 MeV in the center of mass system, has been under study for many years \cite{Buchmann06-NPA}, but is still subject to debate \cite{An15-PRC,deBoer17-RMP}. 

The $Q$ value of this reaction is 7.162\,MeV, and at low energies it is believed to be dominated by capture to the ground state of $^{16}$O \cite{NACRE13-NPA}. At $E$ = 1\,MeV, near the lowest experimental data point available \cite{Fey04-Diss}, the cross section for ground state capture given by a recent evaluation is $\sigma_{\rm GS} \approx 50$\,pb \cite{NACRE13-NPA}, about three times larger than the so-called cascade contributions from capture to excited states in $^{16}$O. The $\gamma$-ray energy is $E_\gamma = Q + E \approx 8.2$\,MeV.

The present detectors HZDR-1 and HZDR-2 have already been used for nuclear astrophysics experiments at the Earth's surface, in typical arrangements with escape suppression  \cite{Schmidt13-PRC,Schmidt14-PRC,Depalo15-PRC,Wagner18-PRC}. Using the $\gamma$-ray efficiency data from Ref. \cite{Wagner18-PRC}, at $E_\gamma$ = 8.2\,MeV the efficiency is 2.7$\times$10$^{-4}$ for HZDR-1 and 2.2$\times$10$^{-4}$ for HZDR-2.

Assuming a running time of 2000 hours, a beam intensity of 50\,\textmu A $^{12}$C$^+$ beam, and a target thickness of 7.5$\times$10$^{17}$ cm$^{-2}$ helium gas, one obtains 12 keV energy loss in the target (in the center of mass system), conservatively rounded up to 20 keV to take into account the  finite $\gamma$-ray energy resolution of the detector. 

Using the known \cite{Wagner18-PRC} detection efficiency and the measured background (Table \ref{Table:8MeV}), assuming the ground-state S-factor from the NACRE II compilation \cite{NACRE13-NPA} and using isotropic $\gamma$-ray angular distribution, one can then predict counting rates from the reaction and from the background: For detector HZDR-1, 21 counts from the reaction and 2 counts background. For detector HZDR-2, 17 counts from the reaction and 1 count background. 

%================================================================================================
\begin{table}[t]
\caption{Observed background counting rate in the $E_\gamma=7.5-8.5$\,MeV energy region, with active muon veto in Felsenkeller, for each of the detectors studied here.}
\label{Table:8MeV}
\center
\resizebox{\columnwidth}{!}{%
\begin{tabular}{ll|c}\hline
Detector & & Counting rate \\
 & & [10$^{-3}$ keV$^{-1}$ h$^{-1}$] \\ \hline
HZDR-1 (90\% HPGe) 				& & 0.052	$\pm$	0.009 \\
HZDR-2 (60\% HPGe) 				& & 0.024	$\pm$	0.005 \\
HZDR-3 (Euroball/Miniball)& add-back & 0.305	$\pm$	0.014 \\
 						& single-sum & 0.319 $\pm$ 0.014 \\ \hline
\end{tabular}												
}% resizebox
\end{table}
%================================================================================================

For detector HZDR-3, only estimates are possible, because no published $\gamma$-efficiency data exist yet. Conservatively assuming that the detection efficiency is 2.7 times greater than for HZDR-1 (240\% relative efficiency for \linebreak HZDR-3, 90\% relative efficiency for HZDR-1), using the 5.9 times larger background (Table \ref{Table:8MeV}), there will still be a signal-to-back\-ground ratio of 4:1, sufficient for highly sensitive experiments. 

It should be noted that these conclusions depend on the cascade scheme of emitted $\gamma$ rays. In the case of the $^{14}$N(p,$\gamma$)$^{15}$O experiment \cite{Wagner18-PRC} used for the present efficiency data, there were mainly double cascades, i.e. two $\gamma$-rays emitted at the same time. In that experiment, the peak detection efficiency was the same with or without the active veto, meaning that the 70\,mm lead collimator protecting the BGO veto from direct $\gamma$-rays from the target was sufficiently thick \cite{Wagner18-PRC}. This precondition must of course be checked for each planned experiment. In the case of the $^{12}$C($\alpha,\gamma$)$^{16}$O reaction, the dominating component, ground state capture, should not be affected, but for the cascade contributions, possible effects of the muon veto on the detection efficiency must be considered.

Summarizing the discussion of HPGe detectors, in credible scenarios a signal-to-back\-ground ratio of better than 10:1 is found for detectors HZDR-1 and HZDR-2, and about 4:1 for detector HZDR-3.

\section{Summary}
\label{sec:Summary}

The $\gamma$-ray background in the new Felsenkeller accelerator laboratory (140 m.w.e.) has been studied in details,  using three HPGe detectors with escape-suppression shields that also serve as muon veto. As in previous studies \cite{Szucs12-EPJA,Szucs15-EPJA}, it was found that the efficiency of the muon veto improves when combined with the passive shield of the rock overburden. The background in the crucial 6-8\,MeV $\gamma$-ray energy region was found to be 500-2400 times lower than in a non-vetoed detector at the Earth's surface. 

The setup and initial background data from a new, very large offline $\gamma$-counting HPGe detector were described. 

Initial tests of the carbon sputter ion source of the new underground accelerator, conducted underground in the new laboratory, have shown stable operations over 6 hours with currents in the tens of \textmu A. 

By combining the $\gamma$-background and beam-intensity data, it is shown that in principle, highly sensitive experiments with HPGe detectors will be possible at the new laboratory, including a study of the $^{12}$C($\alpha$,$\gamma$)$^{16}$O reaction at very low energy.

It remains up to future work to extend the feasibility data to large 4$\pi$ $\gamma$-ray calorimeters in deep \cite{Bemmerer05-EPJA} or shallow underground laboratories.
 
\section*{Acknowledgments}
The authors are grateful to Toralf Döring, Maik Görler, Andreas Hartmann, and Bernd Rimarzig (HZDR) for technical support.
Financial support by Deutsche Forschungsgemeinschaft %\linebreak 
(TU Dresden Institutional Strategy "support the best", INST 269/631-1 FUGG, BE4100/4-1, and ZU123/21-1) and by the Helmholtz Association (NAVI HGF VH-VI-417 and ERC-RA-0016) is gratefully acknowledged.

\bibliography{}

\begin{appendix}
\section*{Appendix}
\renewcommand{\theequation}{A\arabic{equation}}
\setcounter{equation}{0}
\renewcommand{\thetable}{A\arabic{table}}
\setcounter{table}{0}
\renewcommand{\thefigure}{A\arabic{figure}}
\setcounter{figure}{0}

For reference, the background counting rates in detectors HZDR-3 (Table \ref{tab:MB}) and HZDR-2 (Table \ref{tab:60HPGe}) are given in the appendix. For the latter detector, the background at three different locations in Felsenkeller tunnel IV was also investigated: WS is an unshielded workshop room, MK2 is a chamber with heavy shielding made of old steel, lead, and iron, and MK1 is another chamber shielded by serpentinite rock. Despite the very different local shielding thickness and materials, the count rates in this high-energy range are comparable, due to the almost equal \cite{Ludwig19-APP} rock overburden.

%================================================================================================
\begin{table*}[h]
\caption{Detector HZDR-3 (Euroball/Miniball): Recorded counting rates in different energy regions, in units of 10$^{-3}$ counts / (keV h).}
\label{tab:MB}
\center
\begin{tabular}{l l|r@{\,}c@{\,}l r@{\,}c@{\,}l r@{\,}c@{\,}l}
\noalign{\smallskip}\hline\noalign{\smallskip}
& Site  					&	\multicolumn{3}{c}{6--8\,MeV}	&	\multicolumn{3}{c}{8--10\,MeV}	&	\multicolumn{3}{c}{10--15\,MeV}	\\
\noalign{\smallskip}\hline\noalign{\smallskip}
 \multirow{4}{2cm}{add-back mode}	& HZDR, no veto &	292.4	&	$\pm$	&	1.4		&	193.2	&	$\pm$	&	1.2		&	138.4	&	$\pm$	&	0.6		\\
&HZDR, with veto			&	21.1	&	$\pm$	&	0.4		&	7.9		&	$\pm$	&	0.2		&	2.89	&	$\pm$	&	0.09	\\
&Felsenkeller, no veto	&5.94	&	$\pm$	&	0.04	&	4.29	&	$\pm$	&	0.04	&	3.19	&	$\pm$	&	0.02	\\
&Felsenkeller, with veto	&	0.525	&	$\pm$	&	0.013	&	0.188	&	$\pm$	&	0.008	&	0.054	&	$\pm$	&	0.003	\\
\noalign{\smallskip}\hline\noalign{\smallskip}
\multirow{4}{2cm}{single-sum mode}& HZDR, no veto	& 	363.6	&	$\pm$	&	1.6		&	259.9	&	$\pm$	&	1.4		&	202.9	&	$\pm$	&	0.8		\\
&HZDR, with veto			&	19.6	&	$\pm$	&	0.4		&	8.0		&	$\pm$	&	0.2		&	3.55	&	$\pm$	&	0.10	\\
&Felsenkeller, no veto	&8.22	&	$\pm$	&	0.05	&	6.29	&	$\pm$	&	0.04	&	4.90	&	$\pm$	&	0.02	\\
&Felsenkeller, with veto	&	0.528	&	$\pm$	&	0.013	&	0.214	&	$\pm$	&	0.008	&	0.085	&	$\pm$	&	0.003	\\
\noalign{\smallskip}\hline											
\end{tabular}												
\end{table*}
%================================================================================================

%================================================================================================
\begin{table*}[t]
\caption{Detector HZDR-2 (60\%\,HPGe): Recorded counting rates in different energy regions, in units of 10$^{-3}$ counts / (keV h).}
\label{tab:60HPGe}
\center
%\resizebox{0.99\linewidth}{!}{
\begin{tabular}{l|r@{\,}c@{\,}l r@{\,}c@{\,}l r@{\,}c@{\,}l}
\noalign{\smallskip}\hline\noalign{\smallskip}
Site	&	\multicolumn{3}{c}{6--8\,MeV}			&	\multicolumn{3}{c}{8--10\,MeV}			&	\multicolumn{3}{c}{10--15\,MeV}			\\
\noalign{\smallskip}\hline\noalign{\smallskip}																					
HZDR, no active shield	&	85.8	&	$\pm$	&	0.3		&	62.9	&	$\pm$	&	0.3		&	50.86	&	$\pm$	&	0.17	\\
HZDR, with veto			&	3.09	&	$\pm$	&	0.07	&	1.12	&	$\pm$	&	0.04	&	0.479	&	$\pm$	&	0.016	\\
Felsenkeller IV WS, no veto	&	3.82	&	$\pm$	&	0.13	&	2.72	&	$\pm$	&	0.11	&	2.20	&	$\pm$	&	0.06	\\
Felsenkeller IV, MK2, no veto	&	2.73	&	$\pm$	&	0.09	&	2.08	&	$\pm$	&	0.08	&	1.69	&	$\pm$	&	0.04	\\
Felsenkeller IV, MK1, no veto	&	2.26	&	$\pm$	&	0.05	&	1.84	&	$\pm$	&	0.05	&	1.45	&	$\pm$	&	0.03	\\
Felsenkeller VIII, 111, no veto	&	2.20	&	$\pm$	&	0.04	&	1.70	&	$\pm$	&	0.03	&	1.390	&	$\pm$	&	0.018	\\
Felsenkeller IV, WS, with veto	&	0.16	&	$\pm$	&	0.03	&	0.059	&	$^+_-$	&	$^{0.020}_{0.016}$	&	0.035	&	$\pm$	&	0.008	\\
Felsenkeller IV MK2, with veto	&	0.069	&	$\pm$	&	0.014	&	0.033	&	$^+_-$	&	$^{0.013}_{0.010}$	&	0.011	&	$^+_-$	&	$^{0.005}_{0.004}$	\\
Felsenkeller IV MK1, with veto	&	0.035	&	$\pm$	&	0.006	&	0.014	&	$^+_-$	&	$^{0.005}_{0.004}$	&	0.0040	&	$^+_-$	&	$^{0.0018}_{0.0013}$	\\
Felsenkeller VIII, 111, with veto	&	0.036	&	$\pm$	&	0.005	&	0.012	&	$\pm$	&	0.003	&	0.0065	&	$\pm$	&	0.0012	\\
\noalign{\smallskip}\hline											
\end{tabular}		
%}										
\end{table*}
%================================================================================================

\end{appendix}

\end{document}